\documentclass[letterpaper,twocolumn,10pt]{article}
\usepackage{usenix}

\usepackage{xspace}
\usepackage{etoolbox}
\usepackage{comment}
\usepackage{booktabs}
\usepackage{graphicx}
\usepackage{makecell}
\usepackage{multirow}
\usepackage{subcaption}
\usepackage{tikz}
\usepackage{caption}
\usepackage{flushend}
\usepackage[shortcuts]{extdash}
\usepackage{balance}
\usepackage{paralist}
\usepackage{siunitx}
\usepackage{enumitem}
\usepackage{amssymb}
\usepackage{tabularx}
\usepackage{multirow, makecell}
\usepackage{pifont}
\usepackage[linesnumbered,ruled,vlined]{algorithm2e}
\usepackage{cleveref}
\usepackage{hyperref}

\mathchardef\mhyphen="2D 

\newcommand{\eg}{\emph{e.g.,}\xspace}

\newcommand{\ie}{\emph{i.e.,}\xspace}

\newcommand{\sysname}{Junction\xspace}
\newcommand{\jfaasd}{junctiond\xspace}
\newcommand{\Jfaasd}{Junctiond\xspace}
\newcommand{\faas}{FaaS\xspace}
\newcommand{\faasd}{faasd\xspace}
\newcommand{\Kb}{Kernel-bypass\xspace}
\newcommand{\kb}{kernel-bypass\xspace}

\newtoggle{showmarks}
\toggletrue{showmarks}

\iftoggle{showmarks}{
  
  \newcommand{\oldtext}[1]{\leavevmode\textcolor{red}{OLD: #1}}
  
  \newcommand\rf[1]{\textcolor{purple}{~RF:~#1~}}
  \newcommand\ab[1]{\textcolor{teal}{~AB:~#1~}}
  \newcommand\jf[1]{\textcolor{orange}{~JF:~#1~}}
  \newcommand\gic[1]{\textcolor{blue}{~GIC:~#1~}}
  \newcommand\ec[1]{\textcolor{green}{~EC:~#1~}}
  \newcommand\ig[1]{\textcolor{teal}{~IG:~#1~}}
  
}{
  
  \newcommand{\oldtext}[1]{\unskip}
  
  \newcommand\rf[1]{\unskip}
  \newcommand\ab[1]{\unskip}
  \newcommand\jf[1]{\unskip}
  \newcommand\gic[1]{\unskip}
  \newcommand\ec[1]{\unskip}
  \newcommand\ig[1]{\unskip}
}

\newcommand\ssstar{\textsuperscript{$\star$}}
\newcommand\sssec{\textsuperscript{\textsection}}

\begin{document}
\date{}

\title{\Large \bf \Jfaasd: Extending FaaS Runtimes with Kernel-Bypass}

\author{
\makebox[0.95\linewidth][s]{{\rm\hfill Enrique Saurez\sssec\hfill Joshua Fried\hfill Gohar Irfan Chaudhry\hfill Esha Choukse\sssec\hfill}} \\ 
\makebox[0.95\linewidth][s]{{\rm\hfill ~~~Íñigo Goiri\sssec \hfill ~~Sameh Elnikety\ssstar\hfill Adam Belay~~\hfill Rodrigo Fonseca\sssec~~~ \hfill}}\\ 
\makebox[0.9\linewidth][s]{\hfill MIT CSAIL\hfill \sssec{} Azure Research -- Systems \hfill \ssstar{} Microsoft Research\hfill}
} 

\maketitle

\abstract{
This report explores the use of \kb networking in \faas runtimes and demonstrates how using \sysname~\cite{junction24}, a novel kernel-bypass system, as the backend for executing components in \faasd can enhance performance and isolation. \sysname achieves this by reducing network and compute overheads and minimizing interactions with the host operating system. \Jfaasd, the integration of \sysname with \faasd, reduces median and P99 latency by 37.33\% and 63.42\%, respectively, and can handle 10 times more throughput while decreasing latency by $2\times$ at the median and 3.5 times at the tail.
}

\section{Introduction}\label{sec:intro}

\par Serverless is one of the main paradigms for cloud-native programming. It simplifies cloud usage by minimizing operational complexity, allowing fine-grained pricing, and scaling the capacity automatically. The main serverless offering is Function as a Service (\faas). With \faas, the customer writes the code for functions that are triggered by certain events (\eg HTTP invocations or timers). The platform provider handles resource allocation, request routing, and function execution. The customer does not need to manage the infrastructure and platforms that host the services. The invocations are stateless, and the data related to the requests are stored in external data services within the same cloud provider. Since the cloud controls the execution of \faas, it can optimize it better than most other types of service.

\par One of the main infrastructure components that affect \faas performance is networking~\cite{rfaas}. All \faas invocations involve at least one remote procedure call, and usually more, as multiple software components are involved in routing requests to the process running the functions, including gateways and sidecars. Each network round-trip is in the critical path of an invocation and consumes CPU time that could be better utilized to serve more functions or to improve the throughput of a specific function instance. As such, \faas can greatly benefit from having an efficient network stack for all its infrastructure and platform components.

\par \Kb networking is utilized in data centers to enhance the performance of software services~\cite{demikernel, ix, arrakis, mtcp, caladan, shinjuku}. Bypassing the operating system kernel, this technology reduces the number of layers involved and minimizes expensive context switches that occur when using regular network stacks, thereby improving performance. However, due to the complexity of implementation and additional computational costs involved~\cite{demikernel,junction24}, most \faas platforms do not leverage kernel-bypass networking for general applications~\cite{spright}.

\par \Kb networking requires the dedicated use of resources to poll network queues, as notifications of packet reception are typically unavailable in user space (where \kb code resides). In \faas, the naive use of \kb incurs a significant penalty in the number of resources allocated to polling. This is because one polling core is required per hosted function instance, and commonly used libraries (\eg, DPDK~\cite{dpdk}) cannot be securely shared for polling across different tenant functions. This issue is further exacerbated by the fact that most functions are not frequently invoked \cite{shahrad2020serverless}, resulting in more resources being spent on polling than on performing computations for the functions.

\par Newer \kb systems, like \sysname~\cite{junction24}, bring the performance benefits without the complexities and resources overhead involved.  In this work, we demonstrate how we can utilize these systems to increase throughput and reduce warm end-to-end latency in \faas runtimes. Specifically, we evaluate how \sysname \cite{junction24} can be seamlessly integrated into \faasd as its primary execution runtime, reducing tail latencies by up to 81\% and increasing throughput by up to 10 times, without significantly increasing the number of allocated cores per server.
\section{Background}\label{sec:background}
In this section, we describe the core building components of our prototype. We first describe the basic architecture used by many \faas frameworks, and more concretely the architecture of \faasd. Finally, we discuss \kb networking, and the properties of \sysname that make it a good fit for \faas.

\subsection{FaaS architecture}\label{sec:faas}

\begin{figure}[t]
  \centering
  \includegraphics[width=1.\columnwidth]{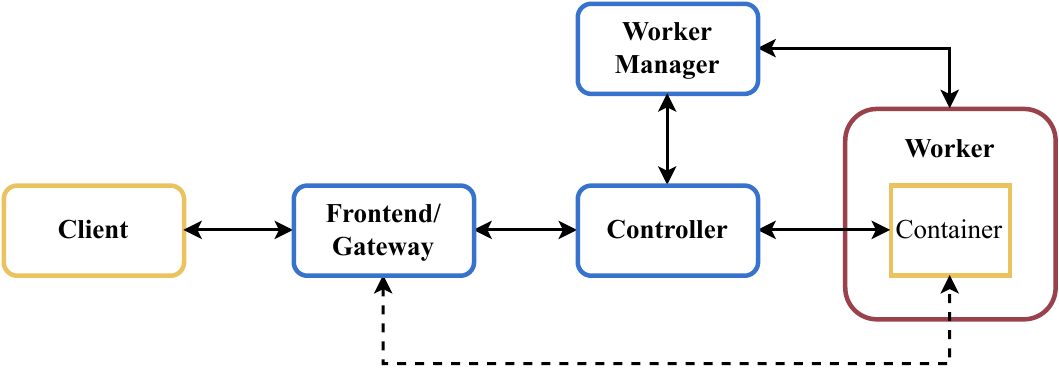}
  \caption{Common FaaS architecture.}
  \label{fig:faas-arch}
\end{figure}

\par \Cref{fig:faas-arch} shows the high-level architecture for the typical \faas platform~\cite{azurefunctions, openwhisk,lambda23,knative,vhive,openfaas}.
A client typically reaches a stateless load-balancer or gateway, which authenticates the request and routes it to the appropriate component. If the function is not currently active, the gateway will request the controller to deploy the corresponding function instance. This operation may also involve adding more workers to the pool via the worker manager if there is insufficient capacity.
\par Once the function instance is ready to handle requests in a worker, the request is forwarded either directly from the gateway or through the controller to the corresponding function instance. Most \faas runtimes execute the function code inside either containers or virtual machines. Additionally, outside of the critical path, the controller will perform auto-scaling operations for both the pool and the function instances to properly handle the load being handled by a given function. 

\par Each of the components in \Cref{fig:faas-arch}, including the gateway, controller, and worker manager, are replicated services deployed on different servers for fault-tolerance. Workers are also typically deployed on separate servers, with their number determined by the overall load of the \faas platform.
 
\par A key aspect of this architecture is that a request must pass through one or more components, such as the gateway, before reaching the container hosting the application. Each additional component in the invocation path adds an additional RPC call and its associated overhead. In some cases, a sidecar may even be present next to the application container to route the request from outside the worker to the process running the function, as it is the case in Kubernetes-based \faas runtimes \cite{knative}.


\subsubsection{faasd}\label{sec:faasd}
\par As a concrete implementation of the architecture presented in \Cref{sec:faas}, we use \faasd as the base building block in our prototype. faasd~\cite{faasd} is an open-source single-node serverless orchestration framework based on OpenFaaS~\cite{openfaas}. 

\begin{figure}[t]
  \centering
  \includegraphics[width=.95\columnwidth]{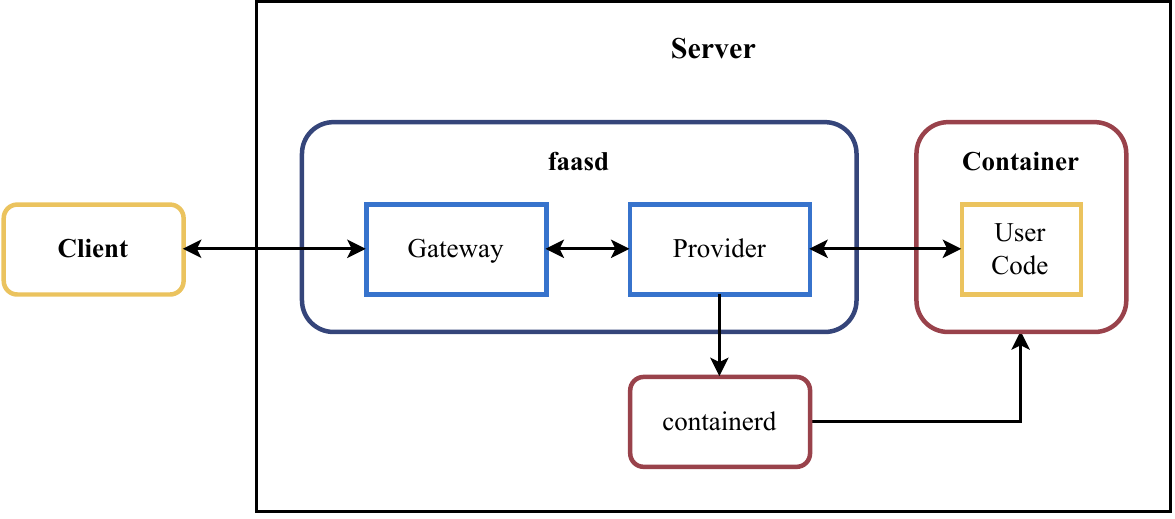}
  \caption{Architecture of faasd.}
  \label{fig:faasd-arch}
\end{figure}

\par As shown in \Cref{fig:faasd-arch},~\faasd employs Linux containers, deployed by containerd~\cite{containerd}, to sandbox untrusted user applications. It includes two orchestration services, both written in Go: a front-end gateway and a provider that communicates with \emph{containerd}. Each of these orchestration services runs as an independent process within the same server.

\par An invocation in \faasd always traverses the gateway and the provider before reaching the container running the user function code. The communication between each of the components is done via gRPC~\cite{grpc}, which means each invocation involves at least three gRPC invocations, plus any additional request to external storage that is common in the context of \faas applications.


\subsection{Kernel-bypass networking}

\par Kernel-bypass networking is a technique that allows user-space applications to communicate with the network hardware directly, without going through the operating system’s network stack. This can improve the throughput and latency of network-intensive applications. Kernel-bypass networking requires specialized hardware and software support, such as network interface cards (NICs) that can access user memory, and user-space libraries that can handle packet processing (\eg DPDK~\cite{dpdk}). 

\par \Kb networking employs polling mode drivers (PMDs) to directly access the network interface cards (NICs), thereby avoiding the overhead of context switching and interrupt processing. PMDs keep a core busy by continuously polling the NIC for incoming or outgoing packets and transferring them between the NIC and user memory. Newer systems are building abstractions atop kernel-bypass libraries to simplify the programming and enable secure multi-tenant usage of the polling resources.
One such system is Junction~\cite{junction24}. which is explained in the next section.

\subsubsection{\sysname}

\begin{figure}[ht]
  \centering
  \includegraphics[width=.95\columnwidth]{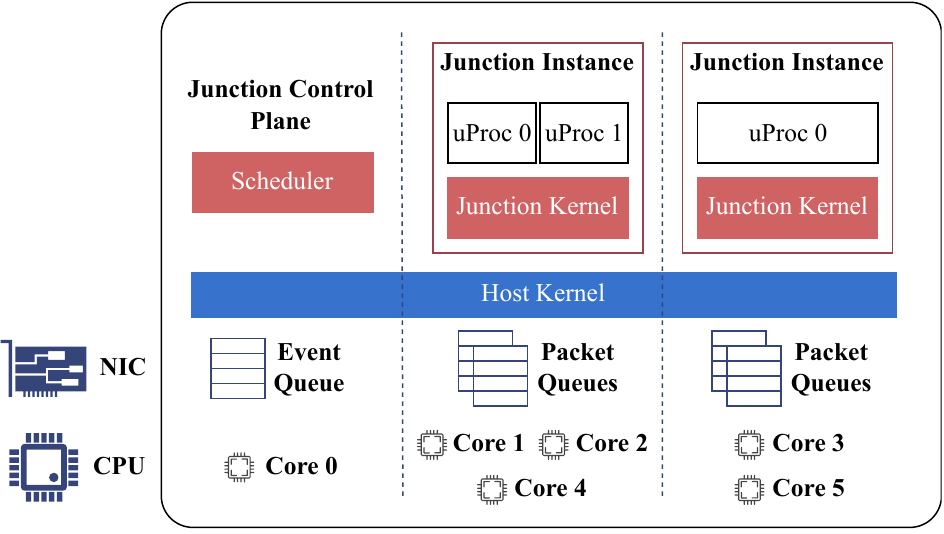}
  \caption{Architecture of \sysname.}
  \label{fig:junction-arch}
\end{figure}

\par \sysname is a libOS-based, kernel-bypass system that can run process-isolated, unmodified Linux applications at a high density, and with practically no performance trade-offs. 
\Cref{fig:junction-arch} illustrates the \sysname architecture.
The three main components in \sysname that are essential for integration with a \faas runtime are:
(1) a \sysname instance, (2) the scheduler, and (3) the I/O management.
For a complete explanation of its architecture, refer to Juction~\cite{junction24}. 

\paragraph{\sysname instance}

\par In \Cref{fig:junction-arch}, a \sysname instance represents a container for executing one or more user applications. Each \sysname instance is a process in the host kernel. 
Each executable within a \sysname instance runs in a user-level, process-like abstraction called a uProc. All uProcs within an instance share the \sysname kernel.
The \sysname kernel provides a Linux syscall abstraction to the uProc,
akin to a library OS ~\cite{exokernel}, enabling it to run existing applications without modifications.

 The use of a libOS-style user-space kernel, in combination with kernel-bypass devices, allows most system calls to be handled entirely within the \sysname instance, except for those necessary to multiplex resources by the host kernel (\ie cores and memory). Shifting OS functionality into user space improves performance by reducing the frequency of context switches and limits the attack surface by allowing untrusted programs to exercise only very small parts of trusted host kernel code.

\paragraph{IO management}

\par The \sysname kernel uses kernel bypass hardware, including both
networking queues and CPU features, to provide its OS
abstractions. Specifically, for the NIC queues, each \sysname instance is assigned one or more NIC send and receive queue pairs, proportional to its maximum core allocation.  To process these queue's packets, the \sysname kernel provides a high-performance network stack. By directly handling the NIC queues in each \sysname instance, it can provide full concurrency across independent instances.    

\par As shown in \Cref{fig:junction-arch}, there are two types of queues that are directly assigned to userspace processes: the packet queues and the event queue. The packet queue, as described in the previous paragraph, is used by instances to communicate with any external entity. The event queue is used to signal when new packets are available in the NIC and is one of the main drivers of the scheduler in \sysname.

\paragraph{Scheduler}
\par The scheduler in \Cref{fig:junction-arch} is used to manage core allocation for all the \sysname instances in a server. The scheduler runs in a reserved core and busy polls on different signals to determine the core allocation for each instance. The allocation of cores varies across time based on demand, up to a configured limit for each instance.  

\par The \sysname kernel provides user-space threads to \sysname instances. This allows the scheduler to have visibility into the state of each instance’s threads (whether they are active or idle) and the signals from the NIC event queues. As a result, the scheduler can centralize polling across all instances. The polling is scalable, as it is mostly proportional to the number of cores active with the instances, rather than being proportional to the total number of instances. This is because the scheduler only needs to make decisions about which task to assign to each core. Additional optimizations are performed to keep the overhead of this decision-making process bounded. The scheduler is also responsible for ensuring fair allocation of cores to each instance and preempting them to assign them to other instances.

\section{Extending FaaS with Kernel-Bypass}\label{sec:design}

\par In this section, we explain how \sysname can be utilized as the backend for executing components in \faasd, thereby improving its performance and isolation. Given the similarity of \faasd's architecture to the general architecture presented in \Cref{sec:faas}, our results can be generalized to other systems.

\begin{figure}[ht]
  \centering
  \includegraphics[width=.95\columnwidth]{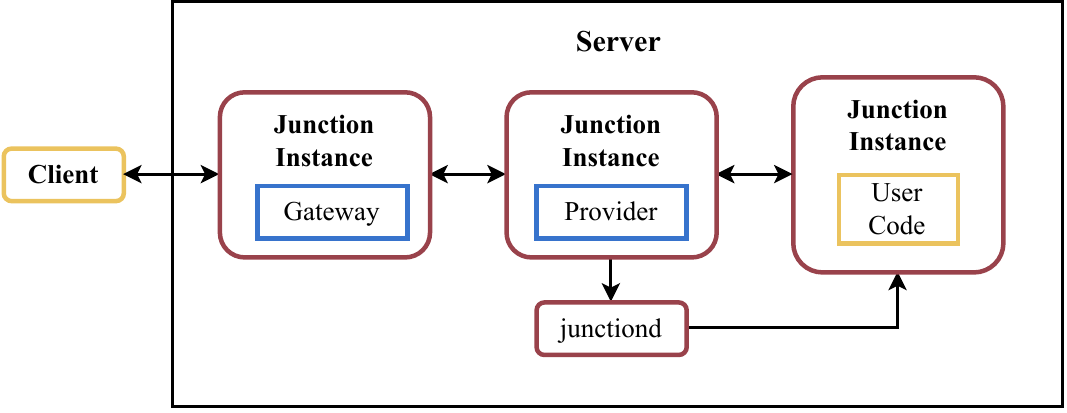}
  \caption{Extending faasd architecture with \jfaasd.}
  \label{fig:jfaasd-arch}
\end{figure}

\par \Cref{fig:jfaasd-arch} shows the mapping of the architecture in \Cref{fig:faasd-arch} to components of \sysname. There are two key changes when using \sysname: a new function manager \jfaasd, and the use of \sysname instances. 

\par \jfaasd serves as a direct replacement for the local container manager, \emph{containerd}. Instead of deploying processes within a container sandbox, they are deployed within a \sysname instance. \jfaasd also manages requests to increase the concurrency of a given function. The scale factor of a function can be modified in two ways, depending on the programming language of its implementation. For language runtimes that do not support native parallelism, such as Python, multiple processes can be deployed within the same \sysname instance. For other languages, the maximum core assignment to a given \emph{uProc} can be modified. If isolation is required across instances of the same function, they can be deployed as independent \sysname instances.
The scale policy decisions are still performed externally by other components in the \faas runtime.

\par \sysname instances are utilized not only to host the function code, but also to run the various services in the \faas runtime (\eg gateway and provider). 
This design choice improves the end-to-end latency and throughput of function invocations, as shown in \Cref{sec:eval}. The use of \sysname instances is well-suited to the overall architecture of a \faas runtime for two key reasons: it provides isolation and increases resource utilization efficiency when compared to using containers and other kernel-bypass systems. 

\par \sysname instances provide greater isolation for function execution compared to containers. This is due to the \sysname kernel’s ability to interpose syscalls and minimize interaction with external components (\ie host kernel), thus reducing the attack surface. Furthermore, Junction delivers packets directly to services and functions through hardware, bypassing the need for software switching. This reduces the likelihood of malicious software accessing it. Given the need for multi-tenancy in \faas runtimes, reducing the amount of trusted code that needs to be reviewed and is vulnerable to attack is highly significant.

\par \sysname{}’s scheduler offers improved resource efficiency compared to standard kernel bypass systems when hosting functions. Typically, kernel-bypass systems require a core to poll for each independent, isolated application instance. However, \sysname’s scheduler’s computational cost is proportional to the number of cores being managed, rather than the number of functions hosted on a server. For instance, \sysname can use a single dedicated core to manage thousands of functions on a 36-core server.
\section{Implementation}\label{sec:implementation}

\par We implemented \jfaasd in C++.  It is a simple component that manages the configuration of junction instances (including network settings), the deployment of instances via the custom `junction\_run' command, and the monitoring the running state of all functions. \Jfaasd is the only component that runs outside of a Junction instance, allowing it to properly spawn isolated junction instances for each function (otherwise the process spawn is handled by the \sysname kernel). Additionally, we implemented a new provider extension that connects the provider process in \faasd to \jfaasd. 

\par We also extended \faasd's provider with a caching mechanism. In mainline \faasd, the provider forwards any state request to \emph{containerd}. However, for our evaluations, we cache these decisions in the provider, assuming that all requests modifying a running function instance will go through \faasd{}’s gateway. This improves the overall \faasd performance, as requests to containerd \emph{can be slower than the function invocation itself and can be on the critical path}. Currently, we only cache the number of active replicas of a function, as well as the associated local IP and port for contacting a function. We use the same caching mechanism with \jfaasd to have a proper comparison.
While it is possible that this caching can improve the performance of \faasd in general, it is beyond the scope of this work to evaluate it
in other contexts.

\par The code is available in GitHub for the benchmark function ~\cite{junctiond_functions}, the modified provider ~\cite{junctiond_provider}, and \jfaasd~\cite{junctiond_manager}.

\section{Evaluation}\label{sec:eval}

\par In this section, we show the benefits in latency and throughput of using \sysname as the execution runtime for \faasd components and for function execution. 

\par \noindent\textbf{Methodology.} This experiment runs on 2 machines with 10 core Xeon 4114 CPUs running at 2.2GHz, 48GB of RAM, and 100GbE NICs. We evaluate the setup using invocations of a serverless function from vSwarm~\cite{vhive,vswarm} that encrypts a 600-byte input with AES. The evaluation compares \faasd with both junctiond and containerd, both with the function metadata caching explained in Section~\ref{sec:design}.

\par \noindent\textbf{Functions benchmark.} To conduct the evaluations, we adapted functions from \emph{vSwarm }~\cite{vhive,vswarm} to work with the templates provided by \faasd. 
For Go functions, we utilized a custom target for the Go compiler, allowing syscalls to be efficiently handled by the \sysname kernel and avoiding the overhead of a trap. For C++ functions, we employed `LD\_PRELOAD' to override glibc with a custom version when loaded into memory, which similarly forwards requests to the \sysname kernel.

\par
\begin{figure}[h]
  \centering
  \includegraphics[width=1\columnwidth]{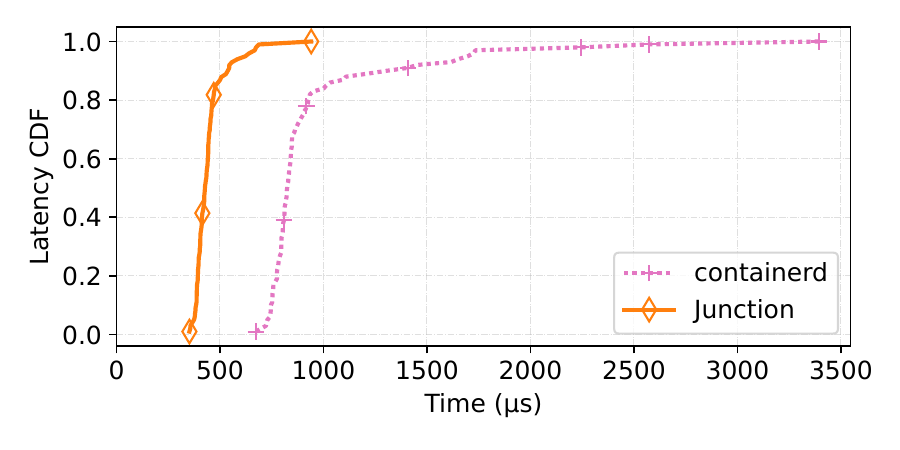}
  \caption{\faasd~\cite{faasd} latency distribution as observed from the gateway for 100 sequential invocations to an AES function~\cite{vhive,vswarm}. \sysname significantly improves the median and tail latency.}
  \label{fig:faasd-cdf}
\end{figure}

\begin{figure}[t]
  \centering
  \includegraphics[width=1\columnwidth]{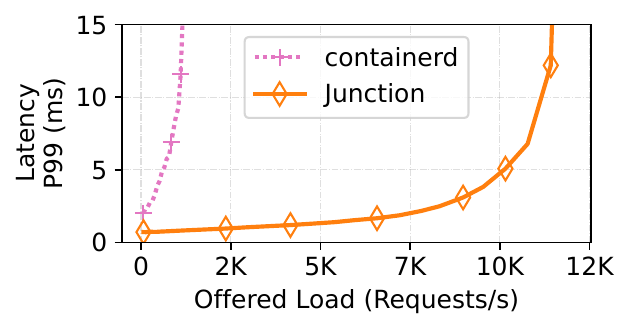}
  \caption{
  \faasd response-time at varying offered loads.
  \sysname offers higher throughput and lower tail latency.
  }
  \label{fig:faasd-load}
\end{figure}

 \noindent\textbf{Average latency.} Figure~\ref{fig:faasd-cdf} shows the latency distribution for 100 sequential invocation to the AES function with 600-byte random inputs. \sysname reduces both median and P99 by 37.33\% and 63.42\%, respectively.  The execution time for the function is also improved, as \sysname performs many of the operating system operations in user-space. The median of the function execution latency is reduced by 35.3\%, while the P99 is reduced by 81\%. \sysname. The improvements shown in both figures \ref{fig:faasd-cdf} and \ref{fig:faasd-load} are due to both the improvements in networking and compute latencies. The compute optimization are related to better thread multiplexing and reduction in context switches due to the \sysname kernel.

\par\noindent\textbf{Tail latency vs load.}
Figure~\ref{fig:faasd-load} shows the tail latency across varying request rates offered via the front-end load balancer.
\sysname can sustain $10\times$ more throughput while lowering the latency by $\sim2\times$ at the median and $\sim3.5\times$ at the tail.
This reflects the compounding end-to-end benefit of using \sysname across multiple components running in separate instances.

\par\noindent\textbf{Cold starts.}
We do not evaluate cold-starts for \jfaasd, but we separately profiled the startup costs for a single-threaded \sysname instance and found that \sysname takes 3.4 ms to initialize them.
\section{Conclusion}\label{sec:conclusion}
We discussed the use of \kb networking in Function as a Service (FaaS) runtimes, specifically using \sysname as the backend for executing components in \faasd.
Our results show that using \sysname can improve performance by reducing latency and increasing throughput, while also reducing the attack surface when compared to containers. \sysname reduces both median and P99 latency by 37.33\% and 63.42\%, respectively, and can sustain $10\times$ more throughput while lowering the median latency by $2\times$ and the tail by $3.5\times$.

\bibliographystyle{plain}
\bibliography{paper}

\end{document}